# Inducing Mimicry Through Auditory Icons

**Hanif Baharin[1*], Norhayati Yusof [2], Suzilah Ismail [2]**

[1]Institute of Visual Informatics, Universiti Kebangsaan Malaysia, 43600 UKM Bangi, Selangor, Malaysia

[2]School of Quantitative Sciences, UUM College of Arts and Sciences, University Utara Malaysia, 06010 UUM Sintok, Kedah, Malaysia

**\* Correspondence:**
Hanif Baharin
hbaharin@ukm.edu.my

**Abstract**

This study aims to find out if periodic auditory icon loop and non-periodic auditory icon loop can induce mimicry in humans. Auditory icons are snippet of everyday sounds used to represent information or processes. A within-subject, Oz-of-Wizard experiment was conducted among forty participants. The participants were asked to eat an apple while being exposed to different types of auditory icon loop. The loops were made using an auditory icon that plays the sound of crunchy apple bite. Both male and female participants were exposed to periodic auditory icon loop, with the auditory icon played every 10 second. Participants were also exposed to non-periodic auditory icon loop which uses the same auditory icon but was made to represent the eating behaviour of a real person of the same sex. The results show that only male participants mimicked the male non-periodic auditory icon loop. Although female participants mimicked the female auditory icon loop, the result is not significant. Both male and female did not mimic the periodic auditory icon loop. Thus, only auditory icons that represent normal biting pace can induce mimicry, significantly in male participants. The findings from this study has implications on the design of persuasive technology that uses auditory icons to encourage behavioural change.

**Introduction**

This paper reports a study which sets out to explore if mimicry of eating behaviour can be induced by periodic and non-periodic auditory icon loops. In this study, mimicry is defined in the scope of human social interactions as 'doing what the others are doing' (1) or also known as the 'chameleon effect' in which people unconsciously copy the verbal or non-verbal behaviours of their interactional partners (2). Mimicry is behavioural matching which does not occur in time; the coordination of similar or dissimilar actions among humans which occur in-time is known as behavioural synchrony (3). There is an abundance of empirical evidence that shows inter-personal coordination between individuals occurs unintentionally (4). Humans have the ability for behavioural synchrony by giving attention to periodic auditory stimulus (5). This tendency for behavioural synchrony is known as entrainment.

In this study, the idea to test if the use of periodic sounds may influence mimicry is inspired by how humans experience their soundscapes. Soundscape is coined by Schafer to denote the all-encompassing and always-present sounds in the environment (6). Sounds in the natural world, especially sounds made by animals such as birds, follows periodic patterns, either daily or seasonally, and these sounds may be classified as "the rhythm of nature" (7). Studies of soundscape in domestic settings shows that humans' activities also contribute to this periodic acoustic patterns, and that people are not only aware of them (8) but they are also aware when the sounds that they associated with certain time are missing (9). This paper does not claim that the ability to recognise the missing periodic sounds in domestic soundscapes stems from the same mechanism as that in knowing the missing beat in musical rhythms. Thus, it is important to make a distinction between recognisable periodic sounds in soundscapes and rhythms in music. Rhythms in music can be defined as "the phenomenal pattern of durations (more precisely, interonset intervals or "IOIs") and dynamic accents." (London 2002, p.531). In music theory, meter is viewed as either a constituent of rhythm or being distinct from rhythm. Meter is "a stable, and recurring patterns of temporal expectations" (London 2002, p.531). People can perceive the interval between beats in music if it is between 100ms to 2000ms (10), and entrainment to the beats means that they know when to expect the next beat, even after the beat in music stops playing.

This study uses auditory icons to represent human eating behaviour. Auditory icons are everyday sounds used to represent computer processes and information (11). Auditory icons are used in sonification, or the representation of non-auditory information using sounds (12). Since people are aware of the periodic sounds in their domestic environment, this research argues that periodic environmental sounds in the form of auditory icons may induce mimicry due to humans' tendency to be entrain to periodic sounds. In order to differentiate between rhythm in music and periodic environmental sounds that people experience in their domestic soundscapes, this research used a periodic auditory icon loop that plays every ten seconds, outside the range that people can recognise rhythms in music. If the sound used is played every 2 seconds, it would sound like music beats instead of everyday periodic environmental sounds.

Technology designed with the aim of changing human behaviour and attitude is known as persuasive technology (Fogg 1998). This paper argues that persuasive technology paradigm may use periodic sounds, in the form of auditory icons, to persuade people to mimic healthy eating behaviour, such as eating slowly, which have been shown to reduce the amount of energy intake in one meal (13). Hermans et al. (2012) shows that when two women eat together, they are likely to mimic each other's food intake action, thus behavioural mimicry may play a part in the amount of food intake since it is influence by food intake of eating companions. Physical presence of companions is not necessary for mimicry to occur. Koordeman et al. (2011) demonstrate that young adult males, who drink beer while watching a movie, are more likely than young adult females, to mimic actors' beer sipping action in the movie. Persuasive technology has been applied in health and wellbeing (R. Orji and Moffatt 2016) - especially in promoting physical activity through mobile applications (16), learning (F. A. Orji, Vassileva, and Greer 2018), and sustainability (17). Meanwhile, ambient persuasive technology model argues for the concept of embedding context-aware persuasive technology in the user's social environment. In this model of persuasion, the process of changing people's behaviour is seen as a process between sources that request for certain types of behaviours and the receivers who fulfil the request (18). The model shows mimicry and repetition are among the characteristics that the source should have in order to increase compliance to a request (18). An example of persuasive technology that influence the change of eating behaviour is Mandometer, which is a scale that represents eating speed in graphical form (19). Mandometer has been proven to change eating behaviour that contributes to health benefits (19).

A review of previous research shows that there is a strong evidence that females have higher tendency than males for nonconscious mimicry due to biological differences (20), for example, females are more likely to have higher number of mirror neurons (21), which has been suggested to be responsible for mimicry (22), compared to male. Experiment of behavioural synchrony using same sex dyads shows that female are more likely to synchronise with their female interactional partner (23). Thus, this research paired male and female with auditory loops that represent eating behaviour of their corresponding perceived sex.

As stated earlier, humans have the tendency to entrain to musical rhythms and produce behavioural synchrony, which can include mimicry of behaviours in time. Since people can also recognise periodic sounds in their domestic soundscapes, this study aims to find out if periodic environmental sounds in the form of a periodic auditory icon loop are more likely to induce people to mimic compared to non-periodic auditory icon loop. The results obtain from this study may be helpful in designing ambient auditory persuasive technology. Based on previous research that studies mimicry in same-sex dyads, the hypotheses tested in this study are as follows:

Null Hypothesis 1: Hearing periodic auditory icons that represents apple biting behaviour of a remote person while eating an apple will not result in mimicry of eating behaviour in female human participants.

Hypothesis 1: Hearing periodic auditory icons that represents apple biting behaviour of a remote person while eating an apple will result in mimicry of eating behaviour in female human participants.

Null Hypothesis 2: Hearing periodic auditory icons that represents apple biting behaviour of a remote person while eating an apple will not result in mimicry of eating behaviour in male human participants.

Hypothesis 2: Hearing periodic auditory icons that represents apple biting behaviour of a remote person while eating an apple will result in mimicry of eating behaviour in male human participants.

Null Hypothesis 3: Hearing non-periodic auditory icons that represents apple biting behaviour of a remote female person while eating an apple will not result in mimicry of eating behaviour in female human participants.

Hypothesis 3: Hearing non-periodic auditory icons that represents apple biting behaviour of a remote female person while eating an apple will result in mimicry of eating behaviour in female human participants.

Null Hypothesis 4: Hearing non-periodic auditory icons that represents apple biting behaviour of a remote male person while eating an apple will not result in mimicry of eating behaviour in male human participants.

Hypothesis 4: Hearing non-periodic auditory icons that represents apple biting behaviour of a remote male person while eating an apple will result in mimicry of eating behaviour in male human participants.

**Materials and Methods**

**Participants**

All participants were briefed about the study and signed a consent form. Forty participants (20 females and 20 males, all aged 19 years old), volunteered to be involved in this experiment. The participants were rewarded RM10 (roughly equivalent to USD 2.50) for completing the experiment. The study was deemed as low risk, therefore ethical clearance was not needed and was waived by the Research Management Centre of the Universiti Utara Malaysia according to the university's policy.

**Auditory Icon Loops Production**

The auditory icon used in this study is the crunching sound of biting an apple which is 5.6 seconds long. The sound wave for the auditory icon used is shown in Figure 1 .

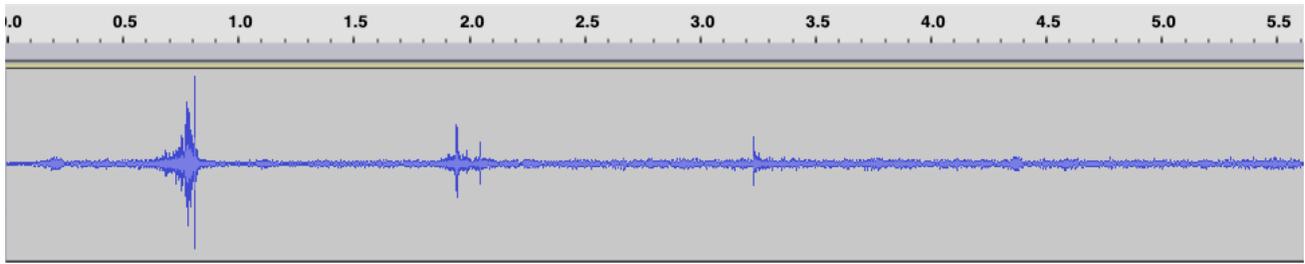

Figure 1: Sound wave of the sound of biting an apple used as the auditory icon in this study

A number of this auditory icon was then combined into a single sound file. The sound file was played in a loop to the participants. This paper will refer to the sound file that contains the auditory icons played in a loop as 'auditory icon loop.' There are three auditory icon loops used in this study – periodic auditory icon loop, non-periodic auditory icon loop for male, and non-periodic auditory icon loop for female.

To create the periodic auditory icon loop, 13 of the auditory icons as shown in Figure 1 are placed ten seconds apart to produce a sound file that is two minutes and six seconds long. The sound wave for the sound file of periodic auditory icon loop is shown in Figure 2.

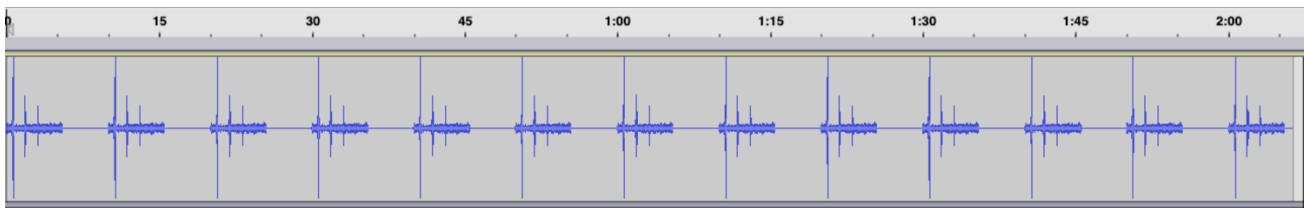

Figure 2: 13 of the apple bite auditory icons are placed 10 seconds apart and combined into a single file to create the periodic auditory icon loop used in this study

To make a non-periodic auditory icon loop, all participants were asked to eat an apple and their eating behaviours were video recorded to keep track of each time they took a bite. From the 20 female participants, one female participant whose eating time is the longest of all the female participants (the slowest eater) was chosen, and from the 20 male participants, one male participant whose eating time is the longest (the slowest eater) of all the male participants was chosen. The slowest eaters from both sexes were chosen because if mimicry occurred it means that it is possible to use auditory icons in persuasive technology to encourage people to eat more slowly. The time of each bite taken by the chosen male and female participants is shown in Supplementary Table 3. The action of one bite of the apple by the chosen participant at a specific time is then represented by one crunching sound of an apple bite (apple bite auditory icon) shown in Figure 1. The resulting sound file for female non-periodic auditory icon loop is shown in the sound wave in Figure 3. The same process was conducted to create the sound file for male non-periodic auditory icon loop (Figure 4).

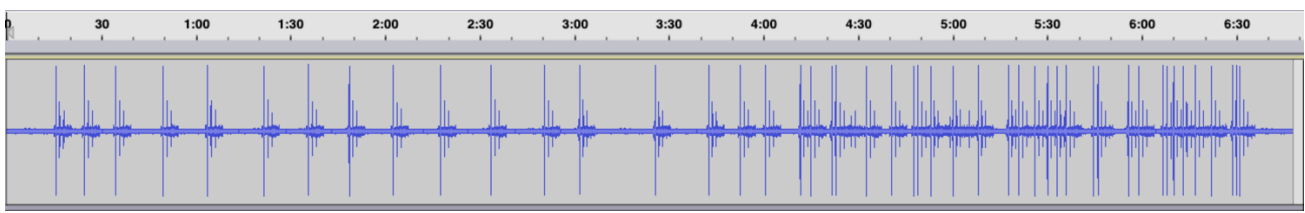

Figure 3: The sound wave for female non-periodic auditory icon loop

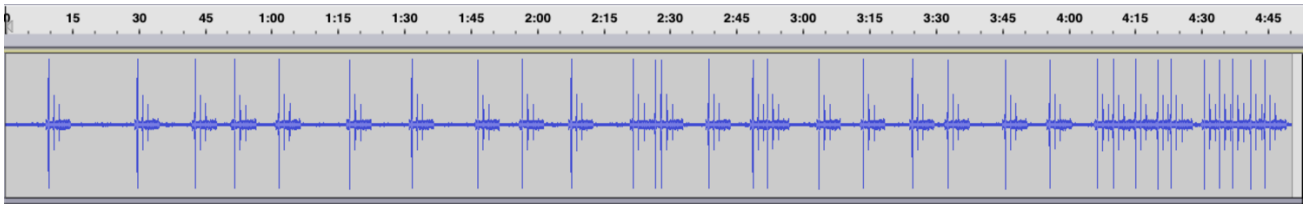

Figure 4: The sound wave for male non-periodic auditory icon loop

**Oz-of-Wizard**

The experiment used a within-subject, Oz-of-Wizard design. Wizard-of-Oz is a technique used in Human-Computer Interaction to study the interaction between users and a low-fidelity software prototype, in which the users interacted with the prototype as if it were a complete and fully functional product. In actuality, the functionalities of the software are simulated by a human on a separate machine (24). Oz-of-Wizard experiment, meanwhile, is an inverse of Wizard-of Oz technique where humans are being simulated by a computer system (25).

**Experimental Procedure**

Based on Oz-of Wizard technique, in the experiments the participants were asked to sit at a table and eat an apple, with a Bluetooth sound speaker on the table that plays auditory icon file as shown above in a loop, while they were eating. The participants were told that there is another person eating an apple remotely at the same time with them, and the sound they could hear from the speaker represents the bite that the person took.

The participants eat an apple in two rounds. Oz-of-Wizard design was implemented where the periodic and non-periodic auditory icon loop were assigned randomly to the participants between these two rounds. Since peridoc auditory icon loop was made by placing the auditory icon 10 seconds apart, there is no male or female peridoc auditory icon loop. Thus all participants were exposed to the same periodic auditory icon loop (Figure 2). Meanwhile, since the experiment's aim is to test mimicry based on same-sex dyads, male participants were exposed to male non-periodic auditory icon loop (Figure 4) and female participants were exposed to female non-periodic auditory icon loop (Figure 3). During each round, participants were video recorded while they eat their apples while listening to the auditory icon loops.

**Analysis Strategy**

The videos were analysed to record mimicked bites, non-mimicked bites, and the number of auditory icons heard by each participant. Since the participants were exposed to auditory icon loops, the number of auditory icons they heard depends on the time they took to finish eating their apples. The longer they took to eat, the longer they will be exposed to the auditory icon loop, therefore the number of each auditory icons they were exposed to also increase. Mimicked bites are defined as the bites the participants took during five seconds after an auditory icon is played (sensitive period). This research chose five seconds as the sensitive period based on a study by Hermans et al. (2012), which looks at mimicry between co-eating participants. According to Hermans et al. (2012), five seconds was chosen to prevent overrepresentation of mimicry because bites during eating occur at a higher pace compared to sipping, as shown in previous research on mimicry of alcoholic drinks intake (15,26) Non-mimicked bites are those taken outside the five second period after an auditory icon is played (non-sensitive period). Total sensitive and non-sensitive periods for each participant were then calculated. Total sensitive period is the number of times the participants heard an auditory icon multiplied by five seconds, while total non-sensitive period can be derived from the total time each participant took to eat the whole apple minus the sensitive period. Finally, the rates of sensitive bites are calculated as follows,

$$R_{sensitive} = \frac{total\ mimicked\ bites}{total\ sensitive\ periods\ (in\ seconds)} \quad (1)$$

Meanwhile, the rates of non-sensitive bites are calculated as follows,

$$R_{non-sensitive} = \frac{total\ non-mimicked\ bites}{total\ non-sensitive\ periods\ (in\ seconds)} \quad (2)$$

Mimicry is defined as the bite that the participants take within the five seconds of hearing the sound of individual auditory icon. The number of times the participants take a bite within this time period is added up and then divided by the sum of length of time of the sensitive period to produce the rates of biting in sensitive period (meaning that biting that mimicked the sound heard).

If the participants took a bite outside five seconds of hearing the bite, it is counted as non-mimicked bite. The number of non-mimicked bites is then totalled up and divided by the sum of length of time outside sensitive period to produce the rates of biting in the non-sensitive period (non-mimicked bites). If the rates of bites in the sensitive period is higher than the rates of bites in the non-sensitive period, it means that the participants are more likely to take a bite in the sensitive period compared to the non-sensitive period. Since mimicry is defined as taking a bite in the five seconds after hearing an auditory icons (sensitive period), then we can conclude that mimicry has occurred.

Table 1 shows demonstrated how the calculation for rates of non-sensitive bites and rates for sensitive bites are calculated based on data from a female participant being exposed to female non-periodic auditory icon loop.

**Table 1. The calculation of sensitive and non-sensitive rates.**

| | |
|---|---|
| Participant ID (female exposed to female non-periodic auditory icon loop) | 101S |
| Number of Bites in Sensitive Period *(a)* | 4 |
| Number of Bites outside Sensitive Period *(b)* | 10 |
| Total Bites *(a+b)* | 14 |
| Number of Auditory Icons the Participant was Exposed to *(c)* | 9 |
| Total Time Participants Eat *(t)* | 133 seconds |
| Total Sensitive Period *(s) = (c x 5 seconds)* | 45 seconds |
| Total Non-Sensitive Period *(ns) = (t – s)* | 88 seconds |
| **Sensitive Ratio *(a/s)*** | **0.08888889** |
| **Non-sensitive Ratio *(b/ns)*** | **0.11363636** |

## Results

Descriptive results of the experiment are shown in Table 2. The outcome indicates that male participants took an average of 2 minutes and 18 seconds with 16 bites to finish eating an apple while listening to periodic auditory icon loop. At the same average number of bites, it is 11 seconds slower than average time of eating when exposed to non-periodic auditory icon loop. As for female participants, the average of number of bites and time eating for both types of sound are almost similar which is 17 bites taken within 2 minutes and 46 seconds. The outcome also shown that both genders have almost an equivalent average number of mimicked bites when listening to periodic auditory icon loop. Changing the sound to non-periodic auditory icon loop, the male participants have 8 mimicked bites which is one extra bite from previous sound, and no change in number of mimicked bites for female participants.

**Table 2. Average of time eating, number of bites, mimicked bites and non-mimicked bites for male and female**

| **Exposure** | **Periodic Auditory Icon Loop** | | **Non-Periodic Auditory Icon Loop** | |
|---|---|---|---|---|
| Gender | Male | Female | Male | Female |
| Time eating (in seconds) | 137.7 | 165.5 | 131.4 | 165.5 |
| Number of bites | 15.6 | 17.3 | 16.0 | 17.0 |
| Number of mimicked bites | 6.8 | 7.0 | 8.0 | 7.0 |
| Number of non-mimicked bites | 8.8 | 10.4 | 8.0 | 10.0 |

Figure 5 shows boxplots of difference between two rates for male and female participants. In order to measure the occurrence of mimicry, the difference between sensitive and non-sensitive rates is calculated for each participant (rate differences). Positive value of rate differences means that sensitive rate is higher than non-sensitive rates. Based on Figure 5(a), the range between the smallest and largest rate differences in female participants is smaller than range of rate differences in male exposed to periodic auditory icon loop. Standard deviation value also indicates similar results. Hence, we can conclude that rate differences for male shows more variation as compared to female participants. Meanwhile, Figure 5(b) suggested that male participants tend to show mimicry more than female where about 75% have positive value of rate differences while only more 50% but less than 75% from female indicate positive values.

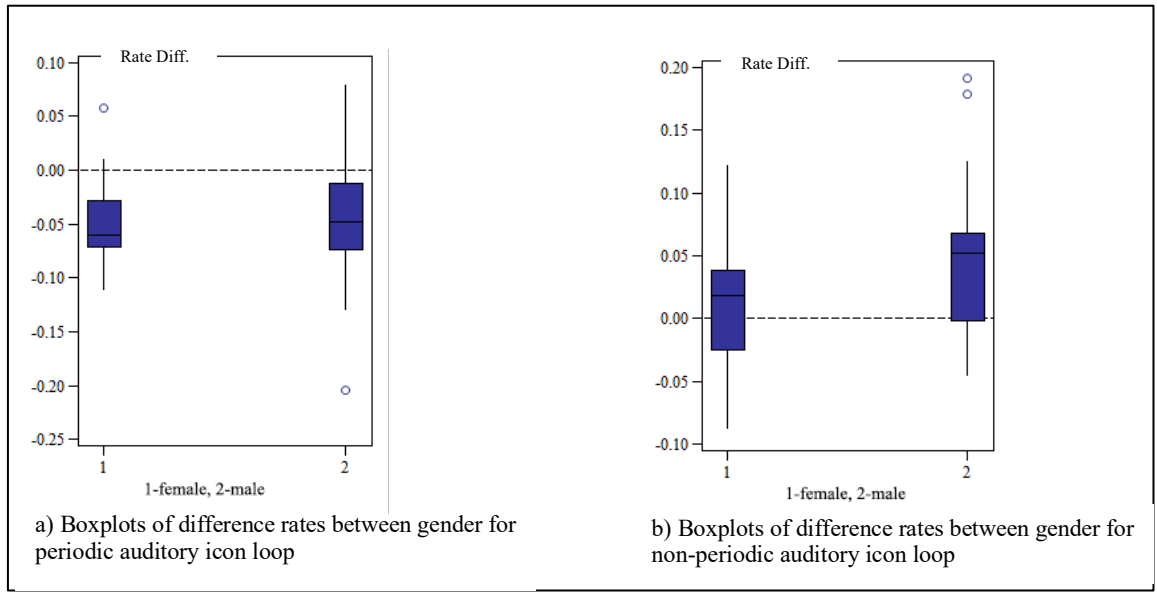

a) Boxplots of difference rates between gender for periodic auditory icon loop

b) Boxplots of difference rates between gender for non-periodic auditory icon loop

**Figure 5. Boxplot of difference rates female and male participants**

A paired sample t-test was conducted to compare the sensitive rates and non-sensitive rates, since both measurements are from each subject (pairs of observations). For the paired samples t-test, the first assumption is independent observations or, more precisely, independent and identically distributed variables and, secondly, the different rates between the two variables must be normally distributed in the population. The first assumption is satisfied since each case (row of data values) holds a distinct person. The second assumption also satisfied after tested using Shapiro-Wilk test.

Table 3 shows that the results are unable to support Hypothesis 1 and Hypothesis 2 although the p-values are small (p<0.0001 for female, p=0.003 for male) and significant at 1%. This is due to the negative value of the average difference rates for both female and male participants who were exposed to periodic auditory icon loop which indicated that the non-sensitive rates are larger than sensitive rates. Therefore, there is no mimicry while eating and hearing the sound of periodic auditory icon loop for male and female. The average difference between sensitive and non-sensitive rates in female participants who were exposed to non-periodic auditory icon loop is positive, indicating that female participants mimic female non-periodic auditory icon loop. However, the p-value is 0.145 which resulted in rejecting Hypothesis 3. Only Hypothesis 4 is supported by this experiment since the average difference rates is positive and significant at 1% (p=0.002) which means male participants mimicked the male non-periodic auditory icon loop with a significance.

Inducing Mimicry Through Auditory IconsTable 3. Summary statistical analysis

| Category | Statistic | Bite Rate while Exposed to Periodic Auditory Icon Loop (number of bites per second) | Bite Rate while Exposed to Non-Periodic Auditory Icon Loop (number of bites per second) |
|---|---|---|---|
| **Female** (*n* = 20) | Minimum | -0.110 | -0.087 |
| | Maximum | 0.058 | 0.122 |
| | Mean | -0.050 | 0.013 |
| | Median | -0.060 | 0.018 |
| | Standard Deviation | 0.041 | 0.054 |
| | *t*-value | -5.448 | 1.091 |
| | *p*-value | <0.0001*** (2-tailed test) | **0.145** (1-tailed test) |
| **Male** (*n* = 20) | Minimum | -0.204 | -0.045 |
| | Maximum | 0.079 | 0.192 |
| | Mean | -0.047 | 0.045 |
| | Median | -0.048 | 0.052 |
| | Standard Deviation | 0.062 | 0.064 |
| | *t*-value | -3.399 | 3.208 |
| | *p*-value | 0.003*** (2-tailed test) | 0.002*** (1-tailed test) |

***Significant at 1%, **Significant at 5%, *Significant at 10%

**Discussion**

The results show that male non-periodic auditory icon loop significantly induce mimicry in male participants. Female non-periodic auditory icon loop did induce mimicry in female participants, but the result is not significant. Since behavioural mimicry is more likely to occur among people with perceived affinity (27), in this study the male and female participants were exposed to the non-periodic auditory icon loop based on their perceived biological sex. The limitation of pairing the participants and their stimulus based on their sex will be discussed in the next section. The results also show that both female and male participants exposed to periodic auditory icon loop, did not mimic the sound.

Periodic sounds occur in nature, made by terrestrial animals based on the day and night cycles, and also the cycle of the seasons. Although this may be termed as natural soundscape rhythm, this study makes a distinction between periodic sounds in environmental soundscapes and rhythms in music. Studies shows that people are aware of the periodic sounds in their domestic soundscapes, and may notice if the environmental periodic sounds are absent during the time when they normally occur (8,9,28). This research speculated that the periodic everyday sounds may have similar effects on humans' tendency to be entrained to rhythmic musical beats. As stated earlier, people can perceive the interval between beat in music if it is between 100ms to 2000ms (10), and entrainment to the beat means that they know when to expect the next beat even after the beat in music stops playing. Drums have been used to produce rhythm to coordinate movement, and hence produce behavioural synchrony in marching, dancing, and also in sports such as the Chinese dragon boat race. Research shows that children as young as 2.5 years old have the ability to synchronize their drum beating to another human playing a drum, much better compared to a machine playing beating a drum using a drumstick or a speaker playing the sound of a drum (29).

Musical tempos also have effects on eating pace. Research shows that people who are exposed to fast tempo music while eating are significantly more likely to eat faster than people who eats while slow tempo music is playing in the background (30,31). Music tempo has also been shown to positively affect the duration of drinking (32). These studies only measured the overall food consumption duration, and not if each individual musical beat will influence people tendency to put food in their mouths. Unlike previous studies that look at music tempo and its effects on food consumption duration, this study aims to find out the effect of periodic environmental sounds on mimicry. In this study, the periodic auditory icon loop is not made to resemble musical rhythm and hence, the auditory icons were place ten seconds apart in the sound file as shown in Figure 2. It is not expected that the participants would be entrained to the periodic auditory icon loop and synchronise their bite to each individual auditory icon played (biting at the same time they hear the sound of an auditory icon from the speaker) but only to see if periodic auditory icon loop can produce mimicry in eating behaviour. However, results show that the participants did not adjust their eating behaviour to mimic the periodic auditory icon loop even though they were told that it represents someone eating in a remote place.

This may be attributed to periodic sound of eating an apple does not represent, generally, the behaviour of humans eating an apple, whole, without cutting it into pieces. Observation of the videos of the experiment shows that the amount of time the participants eat an apple may correspond to the size of bite they took. The larger the size they took for each bite, the longer they needed to chew. Participants who ate like this generally took longer to finish an apple despite taking lesser number of bites. Meanwhile, those who took relatively shorter time to eat, took smaller bites which enabled them to chew faster. It is also most likely that the participants bite size varies throughout the time they were eating to finish the apple. Periodic eating sound, however, represent eating behaviour which may be considered not typical. Taking a bite periodically means that one has to take similar size bite throughout the time of eating and chew at a constant speed. The apple bite auditory icon used in the periodic auditory icon loop are triggered every 10 seconds, meaning that if a person is to mimic the sound, they have to adjust their eating behaviour to take a larger number of bites and chew in a smaller amount of time. Since crunchy apples are quite hard to chew, this did not happen.

Non-periodic auditory icon loops used in the experiment were created based on the eating behaviour of a real female and male participant eating an apple. Therefore, this auditory icon loop resembles a more "normal' way of eating compared to the periodic auditory icon loop, since the pace of eating is determined by bite size and the need to chew. This paper argues that, the participants are more likely to mimic the non-periodic auditory icon loops because they have most probably subconsciously mimicked others' eating behaviour during communal eating thus already attuned to the natural pace of other people's eating. Research shows that syntactic alignment, or mimicry in various aspects of speech, is socially-mediated, and are more likely to occur with people that one perceived to have similarity with (33). There is also a bi-directional relationship between behavioural mimicry and perceived affinity (27). Hence, this paper argues that when hearing a remote participant eating in this 'naturalistic pace' they are more likely to mimic, because this pace is similar to the way the participants eat and they are socially-attuned to this natural pace of eating. The periodic auditory icon loop, however, does not resembles the way the participants eat, and since they are not attuned to this type of eating pace and have no affinity to it, they are less likely to mimic. The reasons why only male participants significantly mimicked the male non-periodic auditory icon loop while the results of female mimicry are not significant is not known and can be explored in future studies.

**Limitation of This Study**

Previous studies are inconclusive about whether sex has an influence on people's social modelling of other's eating behaviour (using other's to guide oneself in food intake), (34). The decision to exposed male and female participants to different non-periodic auditory icon loop that matches their biological sex in this study is based on previous studies that suggest that biologically-sex females are more likely to non-consciously mimic compared to biologically sex males. The decision is also made based on the fact that many studies of eating behaviour use only female participants (34). Despite previous study that suggests that female are more likely to non-consciously mimic their interaction partners, those studies have limitations due the tendency to use same sex dyads without taking into account the socialization of gender, and the difference between a person's identified gender and biological sex (20). Therefore, in the future, experiments will be conducted to exposed male and female participants to the same non-periodic auditory icon loop, and gender and sex data will be collected through self-report, in order to be able to differentiate mimicry between sexes and genders. It must be noted however; the participants were not told about the gender of the remote person the auditory icon loops supposed to represent nor they were asked about their assumptions of the remote partner's sex. This can be taken into account in future studies to find out if perceived sex has an effect on mimicry.

**Conclusion**

Humans have the tendency to mimic many aspects of behaviour during social interactions. They also have the tendency to entrain to a rhythmic external auditory stimulus to produce behavioural synchrony, which is mimicry in time. They are also aware of periodic sounds in their soundscapes and notice if these periodic sounds are missing. Therefore, it was speculated that perhaps people are more likely to mimic periodic environmental sounds much like they tend to entrain their behaviour to mimic the pace of rhythmic musical beats. Based on these premises, the study described in this paper aims to find out if mimicry in eating are more likely to occur when participants are exposed to periodic auditory icon loop compared to non-periodic auditory icon loops that were made based on the real eating behaviour of a male and female participants. The results of the experiment show that the participant's eating behaviour mimicked the non-periodic auditory icon loop but do not mimicked the periodic auditory icon loop. It is argued that people are already mimicking each other while eating together, thus, they are more likely to mimic the non-periodic auditory icon loop that represents natural eating pace. This is in line with other research that shows that affinity positively affects the likelihood to mimic. The results from this study have implications in designing auditory cues that uses environmental sounds such that of auditory icons in ambient persuasive technology.

**Funding**

This study is funded by Universiti Utara Malaysia under the University Research Grant (S/O Code: 13408)

Supplementary Material

**Table 3. Time taken for each bite by the slowest eating female and male participants**

| Time taken for each bite (female) (m:ss) | Time taken for each bite (male) (m:ss) |
|---|---|
| 0:15 | 0:09 |
| 0:24 | 0:29 |
| 0:34 | 0:42 |
| 0:49 | 0:51 |
| 1:03 | 1:01 |
| 1:21 | 1:17 |
| 1:35 | 1:31 |
| 1:48 | 1:46 |
| 2:02 | 1:56 |
| 2:17 | 2:07 |
| 2:33 | 2:21 |
| 2:50 | 2:26 |
| 3:01 | 2:38 |
| 3:25 | 2:49 |
| 3:42 | 2:52 |
| 3:52 | 3:03 |
| 4:00 | 3:13 |
| 4:11 | 3:24 |
| 4:15 | 3:32 |
| 4:21 | 3:45 |
| 4:32 | 3:55 |
| 4:40 | 4:06 |
| 4:47 | 4:10 |
| 4:53 | 4:15 |
| 5:00 | 4:20 |
| 5:08 | 4:23 |
| 5:17 | 4:30 |
| 5:21 | 4:34 |
| 5:26 | 4:37 |
| 5:30 | 4:41 |
| 5:33 | 4:44 |
| 5:36 | 4:50 ( sound file ends) |
| 5:44 | |
| 5:46 | |
| 5:55 | |
| 6:06 | |
| 6:08 | |
| 6:10 | |
| 6:13 | |
| 6:17 | |
| 6:22 | |
| 6:28 | |
| 6:30 | |
| 6:31 | |

6:48 (sound file ends)

# 1   Data Availability Statement

Data will be made online once the paper is published.